\documentclass[preprint]{aastex}
\begin{document}

\title{Diffuse X-ray emission in spiral galaxies}

\author{Krystal Tyler$^{1,2,4}$, A.~C. Quillen$^{1,5}$, Amanda LaPage$^{1,6}$ 
\& George H. Rieke$^{3,7}$}
\altaffiltext{1}{Department of Physics and Astronomy,
University of Rochester, Rochester, NY 14627}
\altaffiltext{2}{Department of Physics, Purdue University, West Lafayette, IN 47907}
\altaffiltext{3}{Steward Observatory, University of Arizona, Tucson, AZ 85721}
\altaffiltext{4}{ktyler@physics.purdue.edu}
\altaffiltext{5}{aquillen@pas.rochester.edu}
\altaffiltext{6}{al007m@mail.rochester.edu}
\altaffiltext{7}{grieke@as.arizona.edu}

\begin{abstract}
We compare the soft diffuse X-ray emission from Chandra images of 12 nearby
intermediate inclination spiral galaxies
to the morphology seen in H$\alpha$, molecular gas, and mid-infrared 
emission.  We find that diffuse X-ray emission is often located along 
spiral arms in the outer parts of spiral galaxies but tends to be 
distributed in a rounder morphology in the center.  
The X-ray morphology in the spiral arms matches that
seen in the mid-infrared or H$\alpha$ and so implies that
the X-ray emission is associated with recent active star formation.

In the spiral arms there is a good
correlation between the level of diffuse X-ray emission
and that in the mid-infrared in different regions.
The correlation between X-ray and mid-IR flux in the galaxy centers 
is less strong.  We also find that
the central X-ray emission tends to be more luminous 
in galaxies with brighter bulges,
suggesting that more than one process is contributing to the level 
of central diffuse X-ray emission.

We see no strong evidence for X-ray emission trailing
the location of high mass star formation in spiral arms.
However, population synthesis 
models predict a high mechanical energy output rate from supernovae for 
a time period that is about 10 times longer than the lifetime of massive 
ionizing
stars, conflicting with the narrow appearance of the arms in X-rays.   
The fraction of supernova energy that goes
into heating the ISM must depend on environment and is probably higher 
near sites of active star formation. 
The X-ray estimated emission measures suggest that the volume filling
factors and scale heights are high in the galaxy centers but low in 
the outer parts of these galaxies.

The differences
between the X-ray properties and morphology in the centers
and outer parts of these galaxies suggest that galactic
fountains operate in outer galaxy disks 
but that winds are primarily driven from galaxy centers.

\end{abstract}


\section{Introduction }

Results from both Einstein and ROSAT have shown that spiral galaxies 
emit soft X-rays ($0.1-2.4$ keV) with luminosities of 
$10^{38} - 10^{40}$ erg s$^{-1}$ \citep{fabbiano,read97,cui}.
Before the launch of Chandra, the study of the diffuse X-ray emission
from spiral
galaxies was seriously impeded by lack of spatial resolution.
Point sources could not be well separated from the diffuse thermal gas, 
making it difficult to compare the morphology of the X-ray
emission to that in other bands.
Study of the morphology of diffuse emission from the disks of spiral galaxies
was therefore possible only for some nearby galaxies
(e.g., 
M101, \citealt{snowden}; 
M51 \citealt{ehle}, NGC~891 \citealt{bregman}).   


The spatial resolution of Chandra not only allows separation between
diffuse emission and point sources, but also allows a more precise
morphological study
of the diffuse emission.  Among normal galaxies, recent Chandra 
imaging has shown
that the diffuse emission can have different types of morphology:
X-ray emitting gas above the plane
of the galaxy that has been interpreted to be caused by winds
escaping the galaxy due to mechanical energy output from massive stars and
supernovae 
(e.g., \citealt{strickland03b,wang01});
coronal X-ray emitting gas associated with
galactic fountains 
(e.g., \citealt{fraternali} in NGC~2403; \citealt{kuntz} in M101).
The diffuse hot gas 
is due to thermal emission from a fairly cool $kT \sim 0.2$ keV 
($T\sim 10^6 K$) component \citep{read97,strickland03}.

The gas detected in X-rays is suspected to be associated with
supernovae and winds from young massive stars.
Since supernovae generate metals, the transport of this hot gas
is key toward understanding the metallicity evolution of galaxies 
(e.g., \citealt{edmunds}).
Superwinds are of cosmological interest as they can transport large 
amounts of gas,
in particular newly synthesized
heavy elements, and energy into the intergalactic medium (IGM).
However, if this gas cannot escape from the disk into
the halo region, it will fall down onto the disk.
In a galactic fountain model,
the hot gas escaping from the disk cools quickly because of 
adiabatic expansion.
Anomalous high velocity HI clouds are suspected
to represent such an infalling phase
of the galactic fountain, whereas the hot
X-ray emitting gas would pertain to the initial outgoing phase of it
\citep{bregman80,deavillez,norman}.  
When the star formation rate is sufficiently high, largescale energetic
outflows are predicted \citep{chevalier,heckman}.  
These ``superwinds'' may account
account for the X-ray emission seen well above the
planes of edge-on starburst galaxies and are potentially capable of
enriching and heating the IGM (e.g., \citealt{martin}).

The morphology and quantity of the hot X-ray emitting gas
is therefore related to the general problem of understanding
the scales (spatial and temporal)
over which metal enrichment takes place in spiral galaxies.
X-ray emission associated with spiral structure should trace
local relatively short timescale galactic fountain type of
metallicity enrichment due to the movement of the spiral density wave, 
whereas larger scale superwinds
are more likely to be relevant to the enrichment of the IGM.

Many Chandra observing programs have focused their studies
on individual galaxies, or on the populations of
unresolved sources in these galaxies.   Only in the last year 
have sufficient number of deep Chandra images of galaxies become available, 
enabling
us to carry out a comparative study of the diffuse components.
The recent study of \citet{strickland03,strickland03b} focuses
on edge-on galaxies.  Here we carry out a complimentary study by focusing 
on low or intermediate inclination galaxies.  These galaxies allow us
to compare the X-ray morphology to the distribution of stellar
components (bulge and spiral arms) and sites of active star formation.

In \S   2 we describe the sample of galaxies and our
procedure for producing images that highlight the diffuse
component.  We also describe comparison data used to 
probe for correlations between galactic components. 
In \S 3 we compare X-ray fluxes and morphology to those observed
at other wavelengths.  A summary and discussion follows in \S 4.

\section{Observations }

\subsection{The Sample}

We searched the Chandra X-ray archive for intermediate or low
inclination nearby galaxies, observed in imaging modes with 
ACIS-S at exposure times that are longer than 20 ks.  
Shorter exposures are not capable of detecting sufficient numbers of photons
from a  diffuse component to enable a good study of its morphology.
We found 12 galaxies that met our criteria; they are listed in 
Table \ref{table:intro} along 
with additional descriptive information.
NGC~1068 and NGC~4258 were omitted from the sample because
of their bright Seyfert nuclei.

\subsection{Diffuse X-ray emission maps}

Reduction of Chandra ACIS-S images was done with CIAO2.3.
Light curves were constructed from the entire S3 chip to allow detection
and removal of background flares.
Because the diffuse emission component is soft we 
removed photons with estimated energies greater than 1.5keV.
This high energy cut increased our ability to detect 
the diffuse component against hard background events, which were 
possibly from low-level flares that were not detected from the lightcurves.  
To ensure that we were justified in making such an energy cut, we made 
color maps of all galaxies with respect to the estimated photon energy.  
The color maps were split into red (0.3 to 1.5 keV), green (1.5 to 2.5 keV), 
and blue (2.5 to 8.0 keV).  The background contained points
of all three colors,
while the diffuse X-ray emission from the galaxies was mostly red, 
confirming that it is soft.  

Point sources were identified from the Chandra event files by masking 
elliptical areas around each source.  
Each region was identified by the CIAO routine
{\it wavdetect} and was then visually inspected to enlarge, if necessary, 
the masking area or to determine if diffuse emission had been misidentified 
as point sources, especially in crowded regions. Pixels inside the ellipses
were filled in based on local background values using the CIAO routine
{\it dmfilth}.  We used the DIST method to fill in the background
because the count rates were not sufficiently high to use the POISSON method.
After point sources were removed we smoothed
the images using the adaptive routine {\it csmooth} with a maximum
smoothing length of 20 pixels.   Images were corrected for the exposure time
and effective telescope area 
using the CIAO routine {\it mergeall} to build the final fluxed images 
which are shown in Figure 1.  
The smoothed images shown in Figure 1 were used for morphological
comparisons.  However, to measure fluxes from particular regions, 
we used the unsmoothed images.

\subsection{Comparison images}

To identify what processes are affecting the distribution
of X-ray gas we searched for images of our sample galaxies
at other wavelengths.  Many of these galaxies were also observed
in CO(1-0) tracing the molecular gas component
as part of the BIMA-SONG survey \citep{helfer}.  
We also downloaded near-infrared H-band images tracing
the stellar component from the 2 Micron All 
Sky Survey (2MASS).
We searched NED for narrow band images of the
H$\alpha$ emission line and mid-infrared images at 12 or 15$\mu$m
from the Infrared Space Observatory (ISO)
that provide tracers of recent star formation.

For NGC~2403, NGC~3184, NGC~6946 and M83 we use H$\alpha$ 
images by \citet{larsen}; for NGC~4579 and NGC~4303 those by 
\citet{koopman}; for NGC~4736, NGC~5055, and M101 those by
\citet{wong}.
Mid-infrared at 12 or 15$\micron$ images 
observed by the instrument ISOCAM are those made available on the NASA
extragalactic database (NED)
by \citet{roussel, bendo}.  For NGC~7331 we downloaded the 15$\micron$
discussed by \citep{smith} from the ISO archive.
The comparison images are shown in Figure 1 with the X-ray images of the 
diffuse emission in each galaxy.

\section{Results}

\subsection{The morphology of the diffuse X-ray emission}

In comparing the diffuse X-ray images with the images of other wavelengths, 
we see that the H$\alpha$ and mid-infrared emission are coincident
each other and with the diffuse X-ray images (see M83 for 
example).  The mid-infrared 
emission is comprised of both continuum and PAH components 
and scales with the rate of star formation \citep{roussel,bendo}, as would 
be expected by dust reprocessing the interstellar radiation field.  The 
images showing the molecular gas (CO) distribution, from which
the stars are formed, also tend to be 
coincident with the diffuse X-ray, though not as closely
as the H$\alpha$ and mid-IR images.

The correlation between the diffuse X-ray morphology
and that seen in H$\alpha$ is consistent with previous studies 
(e.g., \citealt{strickland03, fraternali}). 
However, most previous studies have focused on individual H II regions,
or filaments associated with coronal emission and outflows in edge-on galaxies
(e.g., \citealt{fraternali,strickland03,wang01}).
Because we have focused on spiral galaxies that 
are not at high inclination (except for NGC~7331),
we are better able to view the changes in morphology across the disks of
the galaxies.  We find a clear correspondence 
between the morphology of the diffuse X-ray emission
and the specific locations of active sites of star formation.

Except for NGC~2403, which is irregular,  all the galaxies 
in our sample emit diffuse X-rays from their centers.
Galaxies NGC~3184, NGC~4303, NGC~4736, M51, M83, and M101 show 
spiral structure in the diffuse X-ray emission, coinciding with 
star formation in the spiral arms, as traced in either H$\alpha$
or in the mid-infrared ISOCAM images.  
The emission from the galaxy centers is almost round, whereas
that from the spiral arms is located in linear
features lying along the arms.
Because we see a 
difference in X-ray morphology between the spiral arms and the 
galaxy centers, we compare the two type of regions separately.

\subsection{The diffuse X-ray emission in the spiral arms}

Most but not all of the spiral 
galaxies display X-ray  emission along spiral arms.
Galaxies NGC~3184, NGC~4303, NGC~4736, M51, M83, and M101 show 
prominent spiral structure in the diffuse X-ray emission, coinciding with 
star formation in the spiral arms, as traced by H$\alpha$ or 
mid-infrared emission.  
NGC~4303, NGC~4736, and M101 show weaker spiral structure 
whereas that in in M51, M83, and NGC~3184 is particularly
strong and open (not tightly wound).
Little diffuse X-ray gas was detected in the outer regions of
NGC~6946; however, NGC~6946 has a high Galactic 
absorption column, which may have affected our ability to detect
the soft diffuse X-ray emission. NGC~6946
also has a low star formation rate in its outer disk
compared to the other grand design galaxies in our sample.

Previous studies have shown that the soft X-ray luminosity of entire 
spiral galaxies is proportional to the total infrared luminosity 
\citep{read,strickland03b},    
a direct indicator of the current star formation rate \citep{kennicutt_araa}.
Because we can observe the spatial
distribution of both the diffuse X-ray emission and the current
sites of star formation, we can compare the X-ray and mid-IR fluxes
measured at different regions in each galaxy.
Because longer-wavelength infrared data lacks sufficient spatial
resolution, we opt to use the mid-infrared ISO images.  
However, the total far-infrared
bolometric flux used in previous studies can be estimated from
the mid-infrared flux and is about 14 times the mid-infrared
flux at $12\micron$ (estimated as $\nu F_\nu$) \citep{spinoglio95}.
We have opted to use the mid-infrared ISOCAM images as a tracer of
recent star formation rather than the H$\alpha$ images because
they were observed in a similar fashion  with the same camera
on the same telescope and so are more uniform in their calibration 
than the available H$\alpha$ images. The mid-IR images
also have the advantage that they do not require correction for 
extinction.

In Figure \ref{fig:mir_all} we plot X-ray and mid-infrared
fluxes measured in $20''$  radius
apertures at different locations along the spiral arms of these galaxies.   
Good quality spectra could not be extracted from
the outer parts of these galaxies because of the low count rates,
so we could not determine conversions between fluxes and count rates
based on individual fitted emission models. Consequently we fit spectra
extracted from the brighter (central) components of a few
galaxies and used the resulting model to convert between
counts and flux for other regions.
%
To convert between counts~cm$^{-2}{\rm s}^{-1}$ and unabsorbed 
0.1-2.0 keV flux, we used a {\it XSPEC} MEKAL model with
$kT = 0.45$ keV and an abundance of $0.1$ solar,
which provided a reasonable fit to spectra extracted from
the brighter central components of M51, M83, NGC~5055, NGC~4736 and NGC~7331. 
When Chandra spectra of diffuse emission in nearby galaxies are fit
with single component hot plasma models, they 
have been better fit by those with extremely low abundances.
This problem has been discussed by 
\citet{strickland,wang01} and others.  
Here we do not attach any undue significance to the parameters of
these fits, but use them to derive conversion
factors which we use to estimate physical quantities.
We have noted that the conversion factors are insensitive to the
value of the abundance used.

Background subtracted fluxes were corrected for Galactic absorption based on 
Galactic H I column depth in the direction of each galaxy but   
not for absorption internal to each galaxy.  
The data points used in 
Figure \ref{fig:mir_all} are listed in Table \ref{table:arms}.  

From Figure \ref{fig:mir_all} we see that the X-ray flux in different
regions along the spiral arms of these galaxies 
is correlated with that in the mid-infrared. We find that 
the correlation between X-ray emission and star formation previously seen 
on whole galaxy scales also exists on smaller scales along individual 
spiral arms.  NGC~4579 is the only galaxy with a significantly higher 
X-ray to infrared flux ratio, which is probably due to its AGN 
(discussed recently by \citealt{eracleous}).  
Activity from the AGN has probably increased 
the quantity of diffuse X-ray emission at distances 
3 to 7 kpc from the nucleus.  
Because of this large distance from the nucleus, we infer that
the X-ray gas must have been heated mechanically rather
than radiatively.  Since the surface brightness of the 
diffuse X-ray emission is much lower than 
that seen in the Seyfert galaxies NGC 1068 and NGC 4258, 
we support the conclusions of 
\citet{young} who found that the starburst is {\it not} the dominant source 
of soft extended X-ray emission in Seyfert galaxies.

In the spiral arms of the other galaxies, the flux of X-ray emission is 
about 4 orders of magnitude lower than the mid-infrared emission.   
Using the mid-infrared fluxes to estimate the total infrared 
bolometric fluxes, we estimate that the X-ray fluxes are 5 
orders of magnitude below that comprising the bolometric far-infrared fluxes.  
This is one order of magnitude 
below the correlation found by \citet{read,strickland03b}.  
Part of the discrepancy between our estimated X-ray to
infrared ratio and previous work could be due the
temperature we assumed to convert between counts and flux.
Alternately, the soft X-ray to infrared flux ratio could
also be slightly dependent upon luminosity.

\subsection{Nuclear diffuse X-ray emission}

Except for NGC~2403, which is irregular,  all the galaxies 
in our sample emit diffuse X-rays from their centers.
With the exception of NGC~7331, the nuclei of these galaxies
are also sites of star formation.
NGC~7331 has 
a star forming region ring at about 8 kpc from its nucleus
and little star formation within this ring. However the X-ray
emission peaks at the galaxy nucleus and not at the location
of the molecular ring.
In NGC~7331, the X-ray emission could be associated with 
the bulge of the galaxy,
the star forming ring or its faint Seyfert 2 nucleus.
The bulge in the X-ray emission map north east of the nucleus which resembles
cones seen in other edge-on galaxies (e.g., \citealt{strickland03}), 
suggests that some of the X-ray emitting gas was heated in 
the galaxy center, by the AGN or a past episode of star formation.

We can consider three possible dominant energy sources resulting in
the presence of hot X-ray emitting gas in the centers of these galaxies:
supernovae associated with an old bulge population, superwinds
driven by active star formation, and an active galactic nucleus.  
Since the X-ray emission is not closely associated with individual
spiral arms in the centers of these galaxies, and the 
morphologies are nearly round, a galactic fountain type scenario
is unlikely.  To discriminate
between the possible energy sources, in Figure \ref{fig:x.core} we plot
diffuse X-ray fluxes compared to those in the same apertures  
at H-band (based on the 2MASS images), 
in CO  (based on the BIMA-SONG data) 
and at 12 or 15$\mu$m (based on the ISOCAM images).
Apertures were adjusted to include the entire central 
diffuse X-ray component and have radii which range from 10 to $25''$.
The 0.1-2.0keV unabsorbed X-ray fluxes have 
been measured in the same way as those
along the spiral arms (see previous subsection) 
and are listed in Table \ref{table:core}.

If supernovae associated with older stars were responsible for
the X-ray emission in the nuclear regions of these galaxies
then we would expect a correlation between the nuclear X-ray flux
and that at 1.6 $\mu$m in H-band, tracing the old stellar population.
To test this possibility we show 
in Figure \ref{fig:x.core}a a comparison between the nuclear
X-ray fluxes and the
H-band magnitude measured from the 2MASS images in the same regions.
The X-ray fluxes range over 4 orders of magnitude while the H-band
fluxes range over 2 orders of magnitude.  The range of these numbers
is not strongly affected by the scatter in distances since the 
nearest galaxy in our sample is only 3 times nearer than the farthest one.
We find a correlation between the two fluxes, suggesting that
an old stellar population may be associated with the diffuse X-ray emission.
Either supernovae from an old stellar population contribute to the X-ray 
emitting gas or the mass of the bulge provides a gravitational potential
well that keeps the X-ray gas from escaping the galaxy.  

ROSAT studies of early type galaxies confirmed the correlation between
optical and X-ray luminosity previously revealed by Einstein studies 
(\citealt{brown98}), finding that the X-ray luminosity was
about 4-5 orders of magnitude fainter than the optical luminosity. 
The points in Figure \ref{fig:x.core}a are consistent with this,
suggesting that the diffuse emission in bulges could be an extension
of that seen in early type galaxies.

Figure \ref{fig:x.core}b shows the X-ray and mid-infrared fluxes from 
the galaxy cores in the same apertures as for \ref{fig:x.core}a.  
We see a larger scatter in the X-ray and mid-infrared fluxes in the cores than 
in the spiral arms.  Most of the galaxies are consistent with the 
average mid-infrared to X-ray flux ratio seen in the spiral arms 
(see Figure \ref{fig:mir_all});  
however, there are some outlying points.  
NGC~4579, again, 
has a higher X-ray to emitted infrared flux ratio, which is very likely 
due to the proximity of the AGN.
NGC~6946 has an X-ray/infrared 
flux ratio of $10^{-5}$, two orders of magnitude below the mean seen in 
the spiral arms.  NGC~6946's low level of diffuse central X-ray emission
is either due to its small bulge, or because 
it had been quiescent until recently.
Prior star-forming activity could
affect the total quantity of diffuse nuclear X-ray emission.  

Since we see that the diffuse X-ray flux correlates with the infrared flux 
(and so the star formation rate), it should also be correlated 
with the quantity of molecular gas.  Figure \ref{fig:x.core}b shows X-ray
and CO fluxes for the galaxy cores, using the same apertures.
The CO fluxes were calculated using CO images from the BIMA-SONG survey and
are also listed  in Table \ref{table:core}.  
The fluxes are similarly 
scattered as in a), forming a weak correlation with two outlying points, 
again corresponding to NGC~6946 (upper left) and NGC~4579 (lower right).  
Two of the galaxies in our sample were not included in the BIMA-SONG survey, 
and as such, are not included in this graph:  M83 and NGC~2681.

To test the dependence of diffuse X-ray emission on supernovae from older 
stars and starburst activity, we plotted the ratio of mid-infrared 
to 1.6$\mu$m flux vs. the ratio of X-ray to 1.6$\micron$ flux 
(Figure \ref{fig:x.core}d).  
The lack correlation shown in the plot suggests that current star formation
is not the only factor affecting the quantity of diffuse X-ray flux. 
Unfortunately many
of the high points in Figure \ref{fig:x.core}d are 
also high in Figures \ref{fig:x.core}a,b  
implying that the galaxies with brighter X-ray nuclear
components have bigger bulges and also higher levels
of active star formation near their
centers.  This makes it difficult to discriminate between scenarios
for producing the X-ray gas.
We checked the nuclear spectral classification (listed in Table 1) 
of the galaxies in this plot  and
found no obvious correspondence between nuclear type and X-ray/infrared color.

In short we find weak correlations between central diffuse X-ray flux
and bulge brightness and between the 
X-ray flux and the current rate of star formation.
The correlation between the X-ray flux and mid-IR flux
is weaker than seen in the spiral arms, suggesting that other factors
are affecting the X-ray luminosity.   These other factors could
include the depth of the bulge potential, supernovae from bulge stars,
AGN activity and previous episodes of star formation.

\subsection{Constraints on the volume filling factor, electron density 
and cooling time of the X-ray emitting gas}


As we mentioned above, we used the {\it Sherpa} analysis
program to fit single component hot plasma MEKAL 
models to spectra extracted from the brighter galaxy nuclei.
The MEKAL normalization, 
$K  = {10^{-14}\over 4 \pi D_{cm}^2} \int n_e n_H dV$, required to fit
these spectra, can be used to place constraints on
the emission integral of the X-ray emitting gas 
and consequently on its volume filling factor, $f_v$, and 
electron density $n_e$.
The electron density can be estimated from the 
normalization factor
\begin{equation}
n_e = 0.13 {\rm cm}^{-3} f_v^{-1/2} \left({h \over {\rm kpc}} \right)^{-1/2} 
\left(K \over 10^{-4} A \right)^{1/2}
\end{equation}
by assuming that the emission integral $EI = \int n_e n_H dV \sim n_e^2 f_v h$.
Here $A$ is the aperture area in square arcseconds where we measured
the count rate.
We have scaled the above expressions assuming a typical scale height 
$h \sim 1$ kpc corresponding to $20''$ for a nearby 
galaxy at a distance of 10 Mpc. 
This size scale would be appropriate for the central
components of our galaxies if the central X-ray emission filled
a spherical volume. 

Estimated electron densities in the nuclear regions range from 
$n_e \sim 0.016 {\rm cm}^{-3} f_v^{-1/2} 
\left({h \over {\rm 1 kpc}}\right)^{-1/2}$
in the center of M101 to $n_e \sim 0.11 {\rm cm}^{-3}$ 
(same scaling) in the center of
M83. The irregular galaxy  NGC~2403 has a somewhat lower
$n_e \sim 0.012$ cm$^{-3}$  and NGC~4579 which contains a bright AGN
is higher with $n_e \sim 0.19$ cm$^{-3}$.

Given an estimated electron density and our assumed temperature we
can estimate a radiative cooling time
for the hot gas, 
$t_{cool} \sim 10^7 n_e^{-1} {\rm yr} $ where $n_e$ is in cm$^{-3}$.
This results in a cooling time of 
$t_{cool} \sim 4.5 \times 10^7 f_v^{1/2} 
\left({h \over {\rm 1kpc}}\right)^{1/2} $yr 
for the gas near the center of 
NGC~4579 and $t_{cool} \sim 7 \times 10^8$yr for NGC~2403 and M101.  
These times are quite short, and would be shorter
if the filling factor or scale heights were lower.
This suggests that the X-ray emission from the brighter 
galaxy centers has a fairly high volume filling factor and a large scale 
height, not inconsistent with previous studies
of edge-on galaxies \citep{strickland03}.

Following the same procedures as we previously used for the nuclear 
regions, we can estimate the electron density and cooling time 
for the diffuse X-ray emission in the spiral arms.  The count 
rates in the arms using 20$''$ apertures average around 
10$^{-6}$ cnts cm$^{-2}$s$^{-1}$, corresponding to 
an average electron density 
$n_e \sim 0.006 {\rm cm}^{-3} f_v^{-1/2} 
\left({h \over {\rm 1kpc}}\right)^{-1/2}$,
and an associated cooling time of 
$t_{cool} \sim 1.4 \times 10^9$yr 
$f_v^{1/2} \left({h \over {\rm 1 kpc}}\right)^{1/2}$.

The electron densities in the spiral arms are naturally lower
than those in the galaxy centers.  However, the diffuse X-ray emission 
in the outer parts of the galaxy is mostly located in the spiral arms and 
not in between the arms.  
For the hot gas to be confined to regions near the spiral arms, 
we require that hot gas produced in the arms cool off before it rotates 
away from the spiral arms.  Secondly, less hot gas can be produced interarm.

Hot gas produced in the arms
can cool radiatively or by adiabatic expansion.
In either case, the cooling time must not be longer than
the time it takes for the spiral pattern to travel a large
angle through the galaxy.
For a galaxy rotating at 200 km/s the rotational 
period is 0.3 Gyrs at a radius of 10 kpc from the galaxy nucleus.
Spiral patterns rotate at an angular rotation rate that is approximately
the same size as the angular rotation rate.  
For the X-ray emission to appear in the spiral arms
and be absent between them, the hot gas produced in the arms
must cool faster than the rotational period.
Since our estimated radiative cooling time was much larger than
the angular rotation rate for $f_v \sim 1$
and $h\sim 1$ kpc, we infer that the scale height and 
volume filling factor are probably lower than these values.

We now consider cooling by adiabatic expansion.
If the gas expands by a factor of 3 then the intensity of the
X-ray emission can drop by an order of magnitude and so become
undetectable interarm.
The time for expansion depends on the sound speed $\sim 300$ km/s,
$t_{exp} \sim { 3 f_v h \over 300 {\rm ~km/s}}  \sim 10 {\rm ~Myr} f_v h_{kpc} $.
A time period of 10 Myr would cause the hot gas to lag the
spiral arms.  This suggests that even if cooling took place through 
adiabatic expansion, 
to keep the detected hot gas confined to the spiral arms, 
we would require
a low volume filling factor and scale height $f_v h \lesssim 1$ kpc.

This conclusion is not inconsistent with the narrow 
appearance of the spiral arms, which implies that the hot
gas is not in hydrostatic equilibrium.  Because the ratio of
the sound speed in $10^6$K gas to the rotational velocity is nearly one,
gas in hydrostatic equilibrium would necessarily have $h/r \sim 1$. 
If the volume filling factor is indeed lower than 1, then the 
density in the X-ray emitting medium could be high enough 
$n_e \sim 1$cm$^{-3}$ to correspond to what we might expect from 
X-ray gas originating from supernovae in dense environments.
To be consistent with such an electron density, 
we would infer that $f_v \lesssim 0.01$.

In summary, the high emission measures in the centers of these galaxies
suggest that the volume filling factor and scale heights of the
X-ray emitting gas are high ($h\sim$ a few kpc),
whereas those in the spiral arms suggest the opposite, that the
volume filling factor and scale heights are low 
($h$ less than a few kpc).
It is tempting to consider a picture where
a galactic fountain type scenario operates in the outer parts
of these galaxies, and superwinds are driven from galaxy centers.

\subsection{Comparison of diffuse X-ray emission and star forming regions 
with the predictions of population synthesis models}

We now consider the predictions of population synthesis models.
In the spiral galaxies the star formation rate is highest
just after a spiral arm passes through the ISM. 
H$\alpha$ and mid-infrared emission
peak where the massive young stars are formed.  Population
synthesis models (e.g., \citealt{leitherer95,cervino}) of instantaneous
bursts predict a peak in H$\alpha$ that lasts 
$\sim 3 \times 10^6$ years until the most massive stars leave the 
main sequence.
The energy deposition rate from subsequent supernovae rises at this
time and remains fairly flat until $\sim 3 \times 10^7$ years after
the onset of the burst.  This gives a characteristic timescale
for the expected production of X-ray gas.  Spiral arms contain
H II regions as well as star clusters so clearly there is a range of
ages along the arm (or widths if we think of the spiral arm as a pattern 
passing
through the disk of the galaxy).  So we can't think of the arm
as a wave where at each point we see one and only one 
age after the burst.  However, because H II regions can only last
a few million years and the mechanical energy deposition from 
supernovae takes place
over a timescale 10 times as long, we do expect a difference between
the morphology seen in H$\alpha$ or mid-IR compared to that
seen in X-ray.  In particular we expect that the X-ray emission
will lag or be offset from the H$\alpha$ and mid-infrared emission 
in the spiral arm.

The angular size of the lag depends on the timescale for 
supernovae to occur and the angular rotation rate of the disk
compared to the spiral pattern.   We expect larger angular offsets
or lags in the center of the galaxy where the rotation periods are fast
compared to the outer parts.
At a distance of 5 kpc from the nucleus and with a typical circular velocity
of 200 km/s the rotational period of a star is 150 Myr. The timescale
for supernovae to inject energy is about 30 million years. 
This is one fifth of the rotational period and so 
would correspond to an angular width of 70 degrees.
For radii twice this the corresponding angular width would be 45 degrees,
large enough to be easily detectable in our figures that 
compare the X-ray emission to that seen in 
H$\alpha$ or the mid-IR.

Visually inspecting the images in Figure 1, 
there does not appear to be 
any such lag between the diffuse X-ray and the H$\alpha$ or mid-IR emission 
in any of the images.
The diffuse X-ray emission images have been smoothed to enhance 
the faint diffuse features; however, this should not affect the location 
of the diffuse emission.  When we ran the adaptive smoothing
routine, we restricted the maximum
smoothing scale to be $20''$.   The smoothing would not have moved
the centroid of the X-ray emission, but only degraded the spatial resolution.
We note that we saw no obvious visual evidence for fine structure 
in the diffuse X-ray emission in the non-smoothed maps. 
Higher signal to noise images would
probably reveal more structure in the diffuse component, but 
we still expect that the hot gas
is distributed over fairly large regions.

In Figure \ref{fig:overlay}, we show diffuse 
X-ray contours on top of grayscale images of the H$\alpha$ and 
mid-IR emission for each of the three galaxies M51, M83, and NGC~3184 that 
show strong, open spiral structure in all three wavelengths.  
In M83 and M51 it appears 
that the X-ray emission in some regions extend to somewhat larger radii 
than the mid-IR emission.  However, there is no such extension seen in 
NGC~3184 where the X-ray contours are nearly centered on the regions 
of star formation.

While there is some difference between the distribution of hot
gas and that seen in H$\alpha$, mid-IR, and diffuse X-ray emission, 
there is no clear
lag between the X-ray emission and that tracing massive star formation.
The only major difference is that the X-ray emission is more diffuse
than the H$\alpha$ emission.  This 
would be expected if the X-ray emission arose from
a somewhat larger scale height above the plane of the galaxy than
the H II regions.

Population synthesis models for the X-ray luminosity are also dependent 
upon an efficiency factor $\epsilon$ that represents the fraction of 
mechanical energy that heats the ambient gas to X-ray emitting 
temperatures.  The adiabatic phase for an individual phase depends on 
the density of the ambient medium and would be shorter in a denser medium, 
but the X-ray luminosity could be higher in such a medium because the 
density of shocked material would be higher.
Supernovae that occur in 
denser regions probably produce more X-ray emission \citep{smith_cox}.  
Our observations 
suggest that $\epsilon$ is dependent upon environment and is higher near 
sites of massive star formation than in lower-density inner-arm regions.  
Because the X-ray emission along the spiral arms is quite
narrow, we infer that supernovae from interarm regions must 
produce less soft X-ray emission than those that explode
near the sites of massive star formation.  

\subsection{Comparison of diffuse X-ray emission and star forming regions with galactic fountain models}

The narrowness of the X-ray emission associated with the strong spiral
arms can be used to place a limit on the vertical scale height 
of coronal gas.  The vertical scale height is likely to be smaller
than the width of the features as seen on the sky.  The width
of the features seen in X-ray is typically less than $50''$ wide, 
implying that 
the vertical scale height of the hot gas is likely to be less 
than 1-2 kpc for these galaxies.  
Since this emission is located at 5-10 kpc from
the galaxy centers, this implies that the aspect ratio
of the coronal emission $h/r \lesssim 4$. 
The X-ray gas at large radii in these galaxies is likely to be
in a fairly thin layer.  While the hot gas is likely to be above
the H I disk, it is not likely to be more than a few kpc above the plane
of the galaxy.  We see no evidence for a bright X-ray halo component 
at large radii in these galaxies.  Even in the starburst galaxy M83,
the fact that the X-ray emission is associated with the spiral
arms in the outer part of the galaxy suggests that it is tied
to the disk and not part of the halo. 
This would suggest that a galactic fountain can operate  
at fairly high levels of star formation.

There is however a change in morphology seen in the X-ray emission 
from disk emission associated with spiral
structure in the outer region to a more round distribution in the center
of these galaxies.
The edge-on galaxies atlas by \citet{strickland03}
also suggests that large scale outflows are primarily a phenomenon
restricted to the central few kpc of galaxies.
This suggests that many galaxies undergo a transition at some radius
between low scale height
hot gas disk,  associated with a `galactic fountain'
to a more spherical distribution that might be associated with an outflow.
At smaller radii the angular rotation rate is higher, so star formation
associated with spiral arms will heat the ISM  more frequently.
If the gas cannot cool before the next spiral wave passes by then
each wave additively inputs energy into the hot component
of the ISM.  If this scenario
was correct then the radius at which the galaxy undergoes a transition
in X-ray morphology could depend on the angular rotation rate
and on the cooling time of the hot gas.  Because the angular rotation
rate is dependent upon the mass distribution (and bulge)
this could in part explain
the weak correlation seen between the central X-ray and bulge 
surface brightnesses (Figure \ref{fig:x.core}a).
Alternatively hot gas  from type I supernovae associated with
an old bulge stellar population
could facilitate the driving of a superwind.



\section{Summary and Discussion}

In this paper we have carried out a study of the diffuse
soft X-ray emission in a sample of nearby low and moderate inclination 
spiral galaxies.
By using the high angular resolution possible with Chandra,
we were able to separate between the diffuse component and
that associated with point sources.   We have compared the 
morphology of the diffuse emission with that at other wavelengths,
finding the closest correspondence between the diffuse emission
and indicators of massive star formation such as H$\alpha$ emission
or mid-infrared emission.
In the spiral arms,  the  0.1-2 keV X-ray energy flux is correlated with
that in the mid-infrared tracing recent star formation,
and is about 5 orders of magnitude below that in emitted in the 
far-infrared. 

The outer regions of these galaxies show that X-ray emission is closely 
associated with star formation lying in linear features
following the spiral arms.  Most 
of the galaxies also exhibit diffuse X-ray emission
in the centers; however, the morphology of this component is nearly round.
The level of central X-ray emission is correlated with that
seen in the mid-infrared and that from bulge stars in the near-infrared,
though the correlation between the X-ray and mid-IR 
is not as tight as seen in the spiral arms.  It is likely that more
than one process is contributing to the level of diffuse X-ray
emission in these galaxy centers.

The narrow size of the X-ray emission in the spiral arms 
and estimated emission measures
imply that the volume filling factor 
$f_v \lesssim 0.01$
and the scale height $h$ is less than a few kpc.
Consequently the X-ray gas in the spiral
arms is probably associated with a galactic
fountain and not part of a superwind.   
By a similar argument we reach the opposite conclusion 
for the hot gas in these galaxy centers,
suspecting that the scale heights are larger than a few kpc
and the filling factors are larger.  
The X-ray emitting gas in these galaxy centers could be 
associated with galactic scale winds originating
from the galaxy nucleus.

Contrary to the predictions of simple population synthesis models,
the X-ray emission in the spiral arms does not lag
the mid-IR or H$\alpha$ emission. The narrowness
of the X-ray emission implies that the X-ray gas produced
in the arms must cool off quickly enough that it is not
detected interarm, and that X-ray gas is not efficiently
produced outside of spiral arms.   This suggests that
the fraction of mechanical energy from supernovae that 
heats the ambient gas to X-ray emitting
temperatures is dependent on environment and is higher
in near the sites of massive star formation.


In this paper we have compared the X-ray morphology to tracers of
current star formation.  However, future work could probe the dependence
of X-ray luminosity and morphology on the past star formation history by
comparing these observations to optical and near-infrared spectroscopic
observations.  Future studies could also
probe the relation between the LINER classification and the X-ray properties
of galaxies.

\acknowledgements
We acknowledge helpful discussions with 
Margaret Drenner, Alaina Henry, Don Garnett,
Chip Kobulnicky, Kip Kuntz, Jessica Almog, and Richard Mushotsky.
We are grateful to Tony Wong for providing us with H$\alpha$ images.
We acknowledge support from the REU program NSF grant PHY-0242483,
and Chandra GO grant GO0-1146X.

This research has made use of the NASA/IPAC Extragalactic Database 
(NED) which is operated by the Jet Propulsion Laboratory, 
California Institute of Technology, under contract with the
National Aeronautics and Space Administration.

\begin{figure*}
\plotone{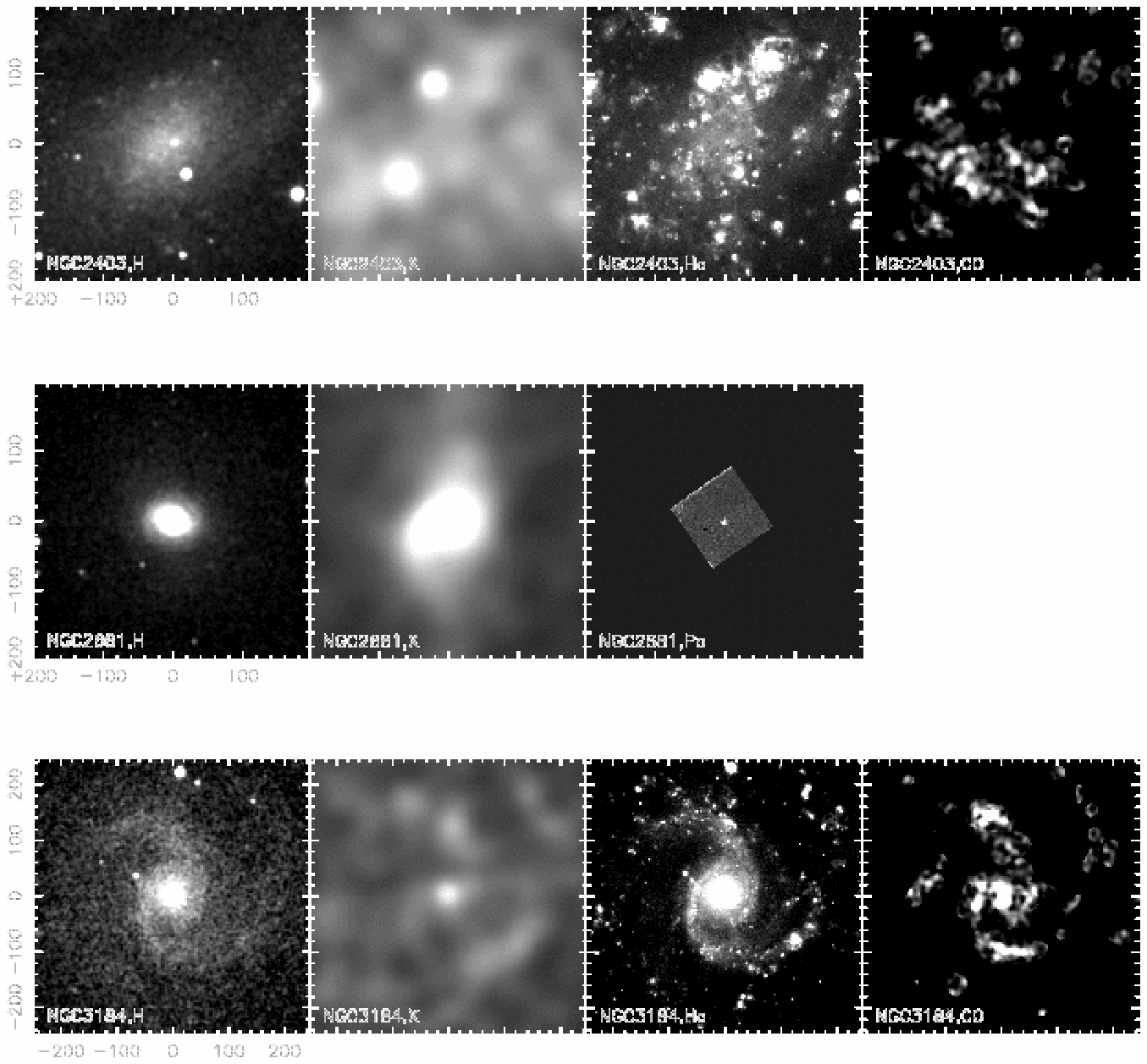}
\caption[]{Diffuse X-ray emission compared to other galactic tracers. 
a) NGC~2403. b) NGC~2681. c) NGC~3814.
From left to right in each panel we show the 2MASS H-band tracing
starlight,
the diffuse X-ray emission from the Chandra ACIS-S image, a narrow band
H$\alpha$ image tracing star formation 
and the molecular gas as seen in CO emission.  
North is up and West is to the right.  Axes are given in arcseconds from
the galaxy nucleus.
A Pa$\alpha$ image is shown instead of H$\alpha$ for NGC~2681.
}
\end{figure*}

\begin{figure*}
\figurenum{1}
\plotone{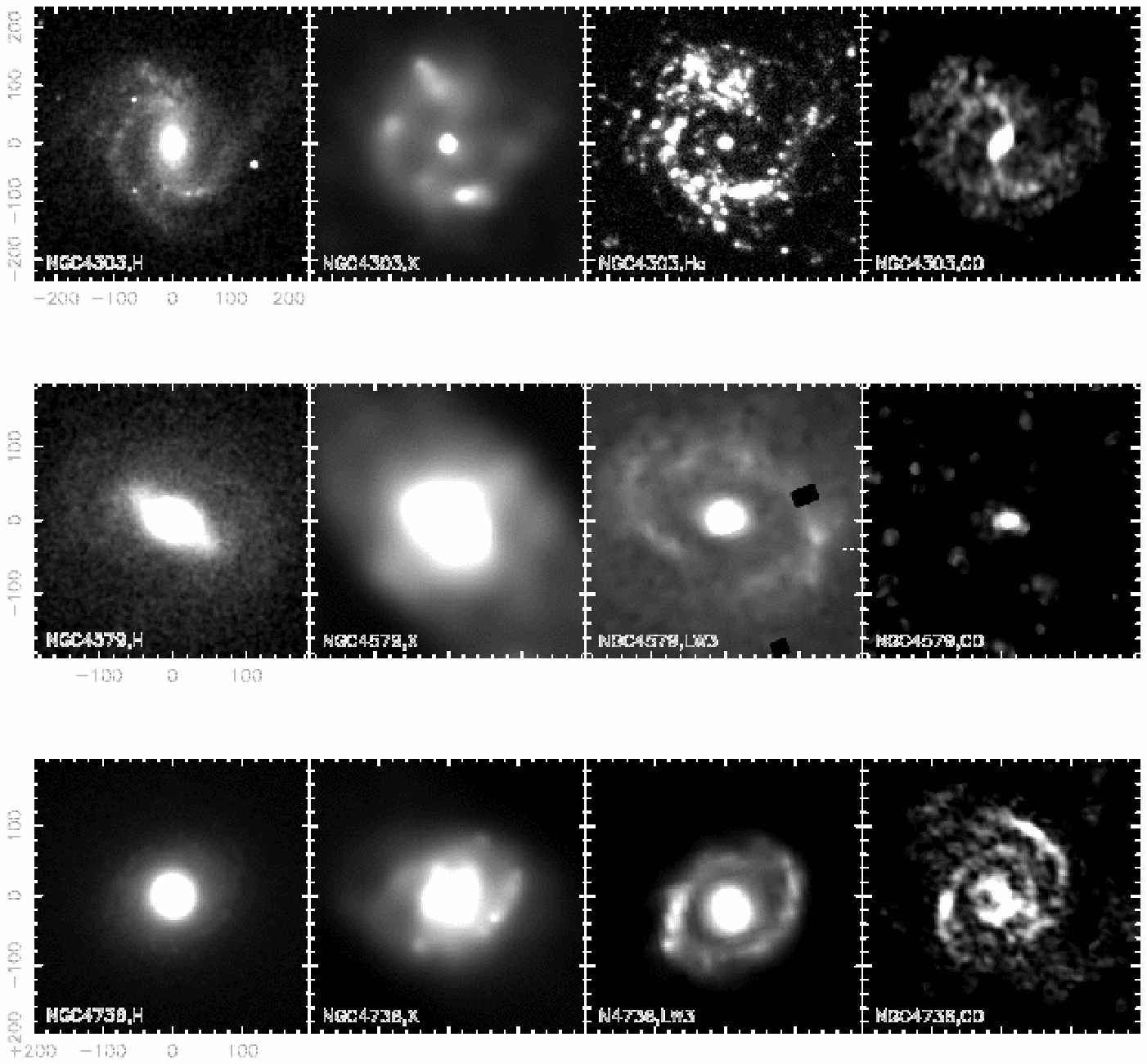}
\caption[]{Continued.
d) Similar to a) but for NGC~4303.
e) For NGC~4579.  f) For NGC~4736.  
The 15$\micron$ LW3 ISOCAM images are 
shown instead of H$\alpha$ for NGC~4579 and NGC~4736 in the third panels.
}
\end{figure*}

\begin{figure*}
\figurenum{1}
\plotone{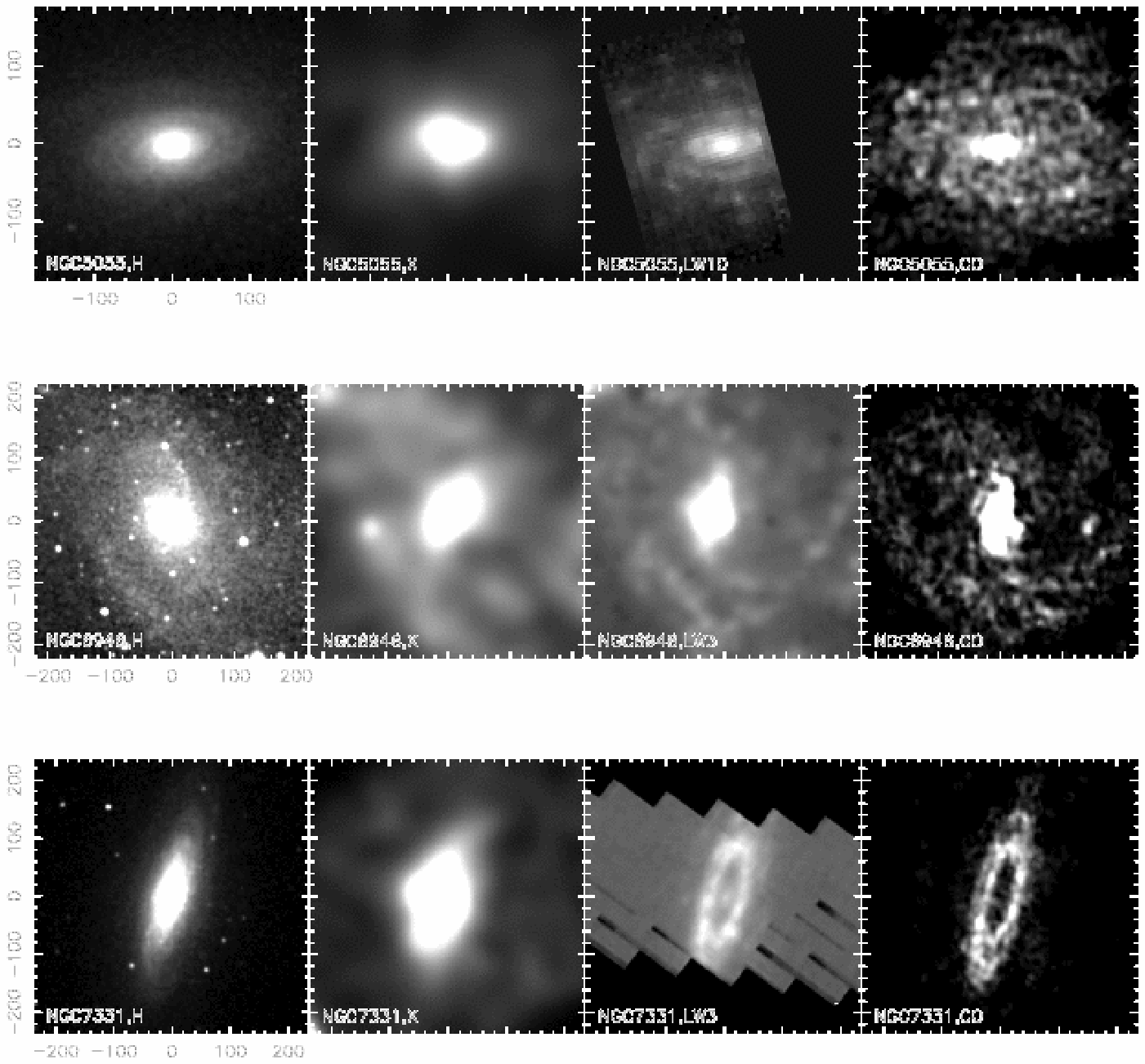}
\caption[]{Continued.
g) Similar to a) but for NGC~5055.
h) For NGC~6946.
i) For NGC~7331.
A 12$\micron$ LW10 ISOCAM image is shown for NGC~5055 and 
15$\micron$ LW3 ISOCAM image are shown for NGC~6946 and NGC 7033
instead of H$\alpha$ in the third panels.
}
\end{figure*}

\begin{figure*}
\figurenum{1}
\plotone{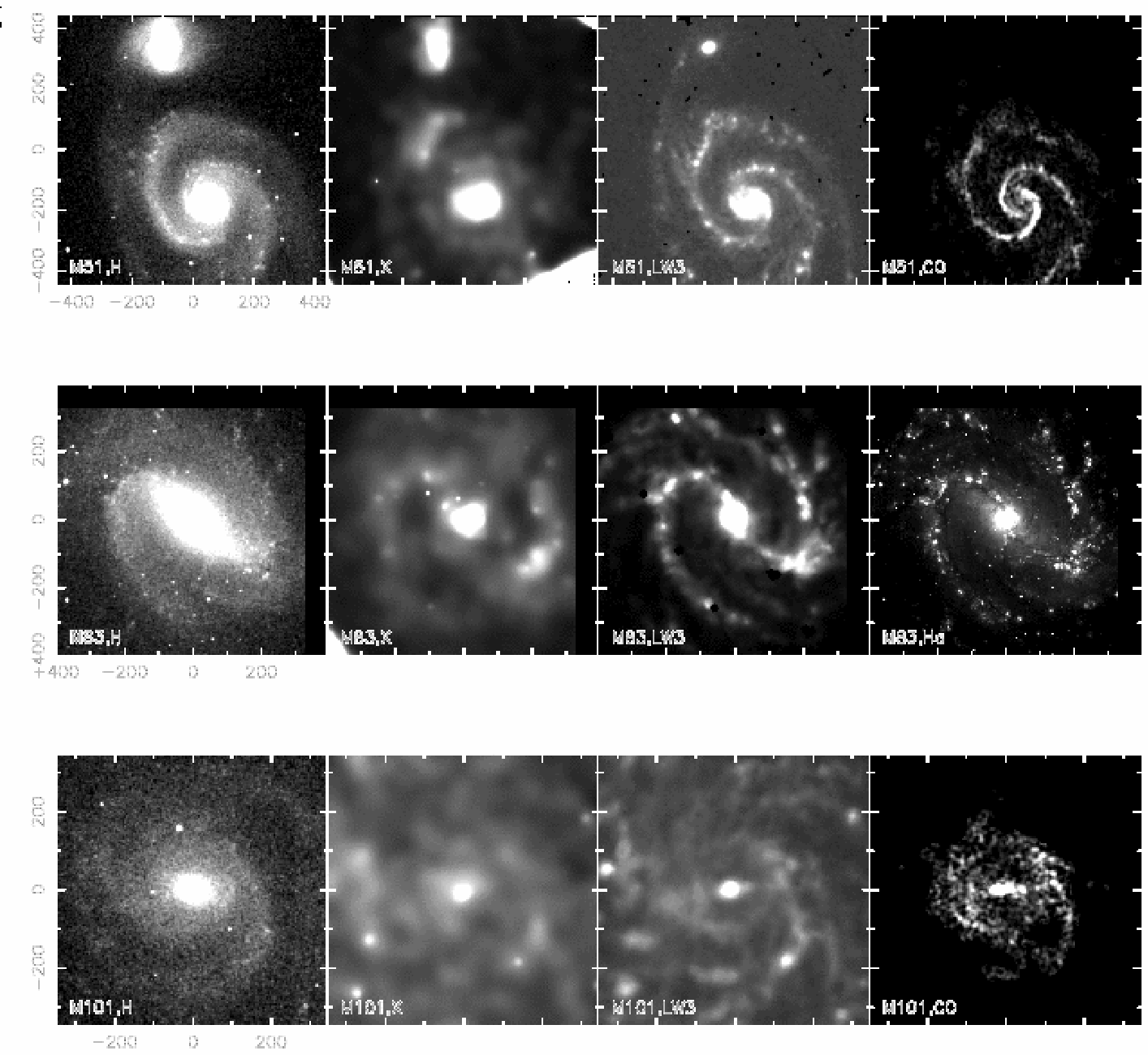}
\caption[]{Continued.
j) Similar to a) but for M51.
Note that the CO image does not cover M51B.
k) For M83.
l) For M101.
The 15$\micron$ LW3 ISOCAM images are 
shown in the third panels instead of H$\alpha$ for M51, M83 and M101. 
Instead of the CO image for M83 we show an H$\alpha$ image in the 4th
panel.
}
\end{figure*}

\begin{figure*}
\epsscale{0.8}
\plotone{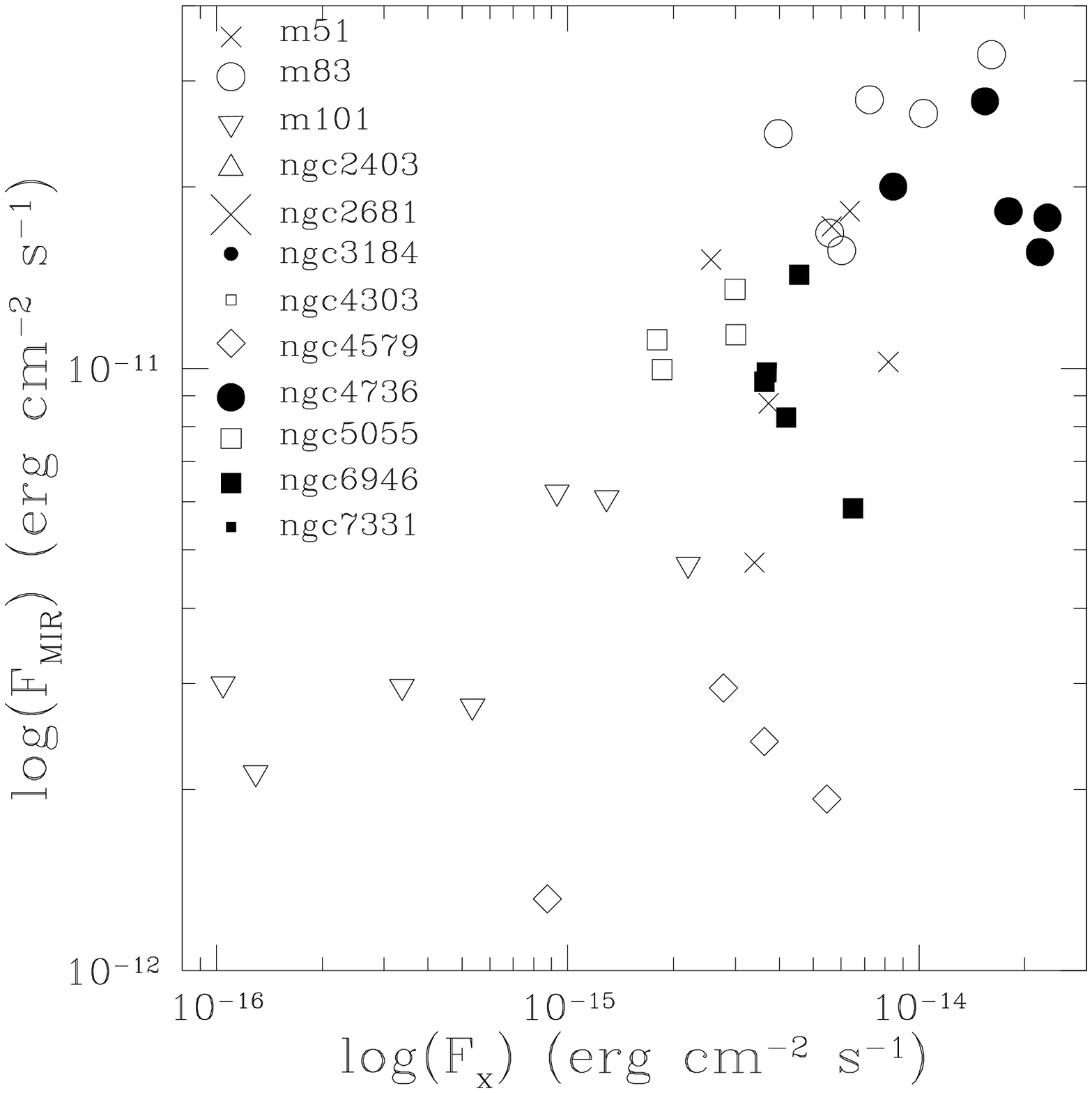}
\caption[]{
A comparison between mid-infrared and X-ray fluxes from $20''$ radii 
apertures in different regions along the spiral arms.  The mid-infrared 
fluxes are measured from the 12 or 15$\micron$ ISOCAM images.
The X-ray fluxes are 0.1-2.0 keV 
unabsorbed fluxes computed assuming a MEKAL
spectrum with $kT=0.45$ keV and corrected for Galactic absorption.
Positions in different galaxies are shown with different shaped data 
points.  The mid-infrared fluxes are about 
3-4 orders of magnitude above 
those seen in the X-rays. The points on the lower right are from NGC~4579, 
which contains an AGN \citep{eracleous}.  
We observe a strong correlation between 
the diffuse X-ray and the mid-IR emission.
\label{fig:mir_all}
}
\end{figure*}

\begin{figure*}
\plottwo{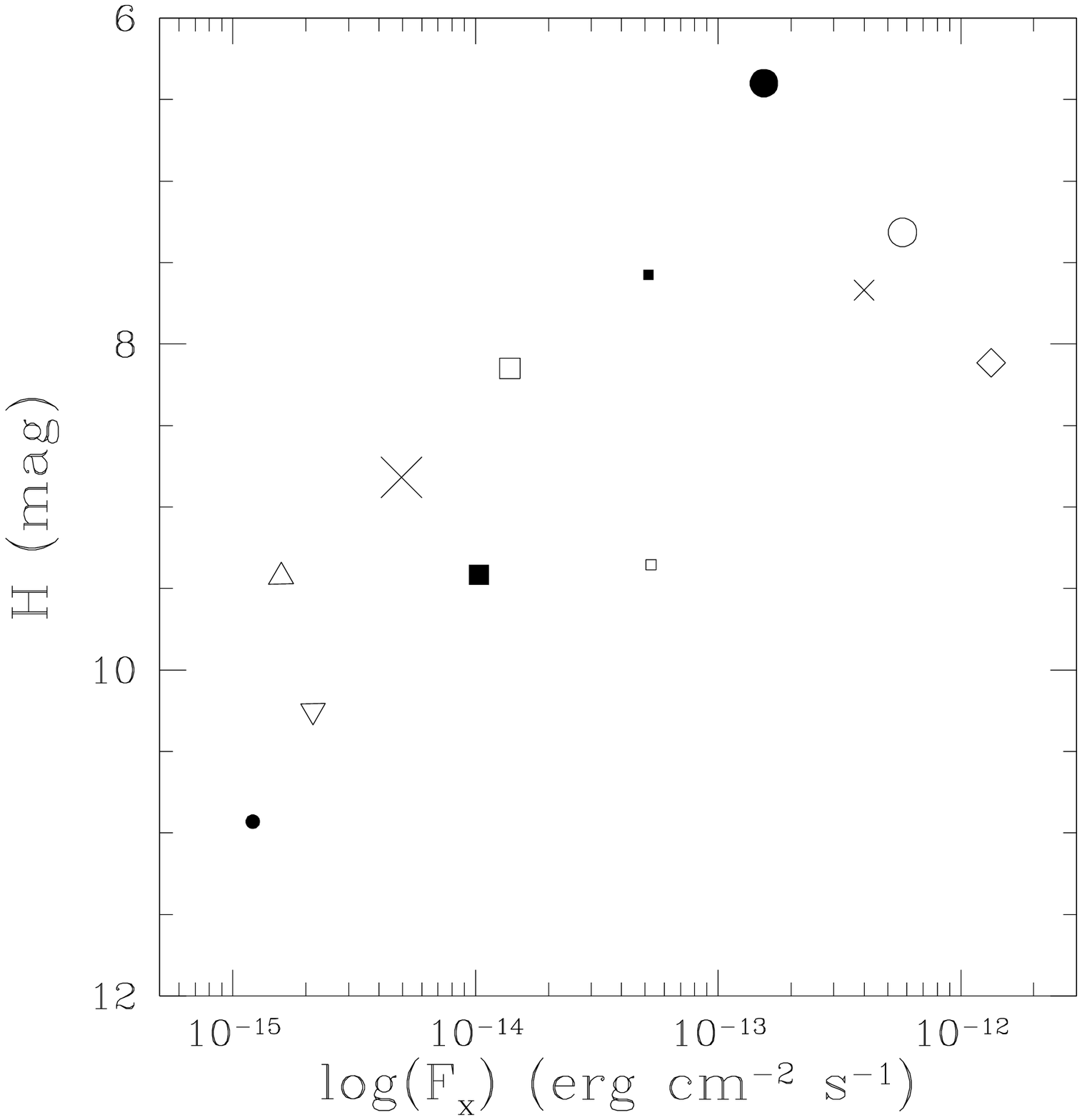}{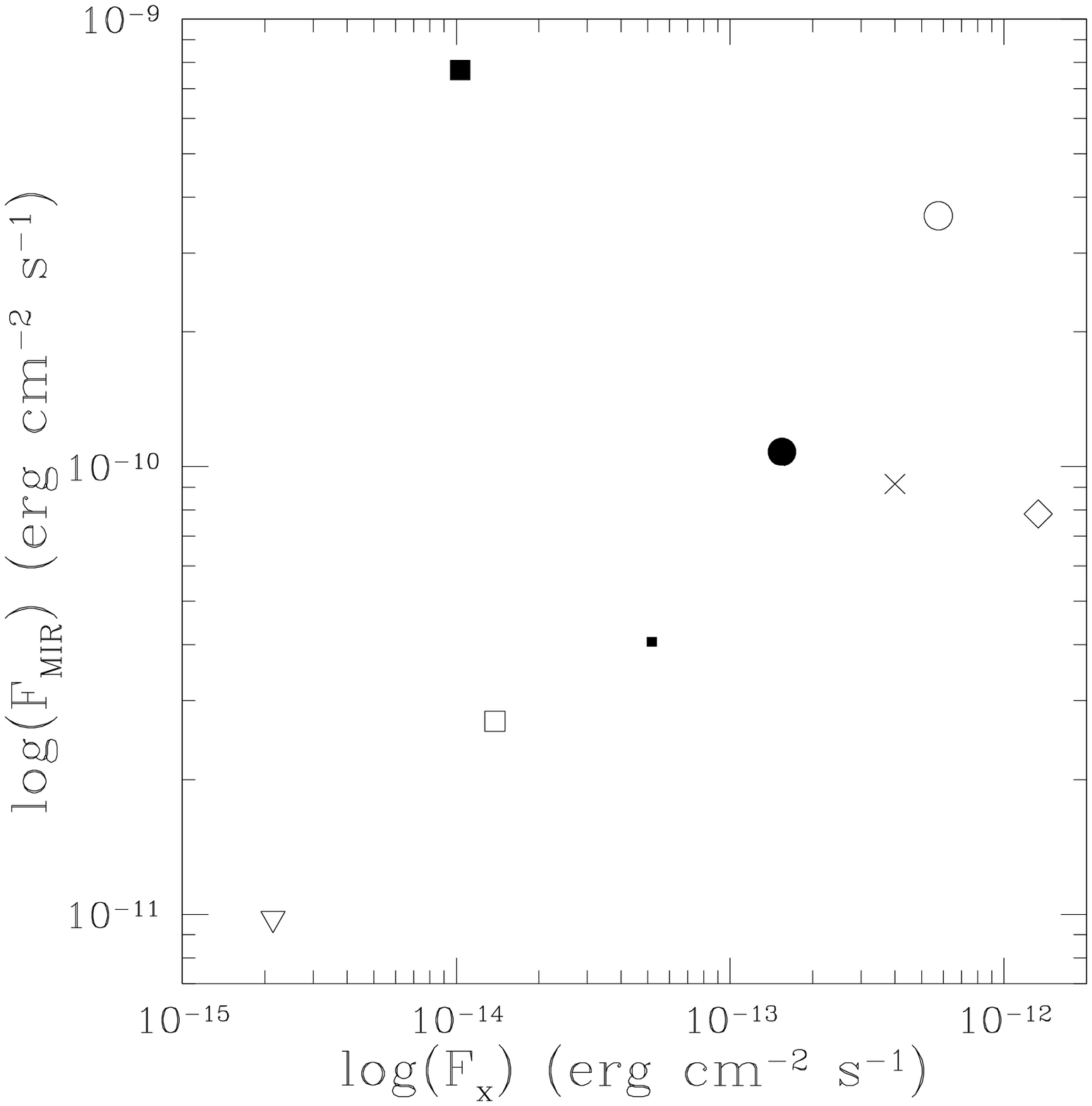}
\caption[]{
Comparison between central diffuse X-ray central fluxes and those
at other wavelengths.
a) We compare X-ray diffuse emission fluxes 
($x$-axis) to the H-band magnitudes ($y$-axis) tracing the stellar bulge
as measured from the 2MASS images in the same apertures.  
Apertures were selected to include the entire nuclear diffuse X-ray emission 
component and range between 10 and $25''$ (listed in Table 2).  
We see a weak correlation 
between the bulge and nuclear diffuse X-ray fluxes.
The data point shapes are the same as shown in Figure \ref{fig:mir_all}.  
b) A comparison between mid-infrared and X-ray fluxes in the cores of 
the galaxies.  We compare X-ray diffuse emission fluxes 
($x$-axis) to fluxes measured from the ISOCAM images ($y$-axis).  
There is more scatter in the 
X-ray to mid-IR core flux ratio than in the spiral arms.  The outlying 
point in the upper left is NGC~6946, while the one at the far lower right is 
NGC~4579, which contains an AGN.
\label{fig:x.core}
}
\end{figure*}

\clearpage 

\begin{figure*}
\figurenum{3}
\plottwo{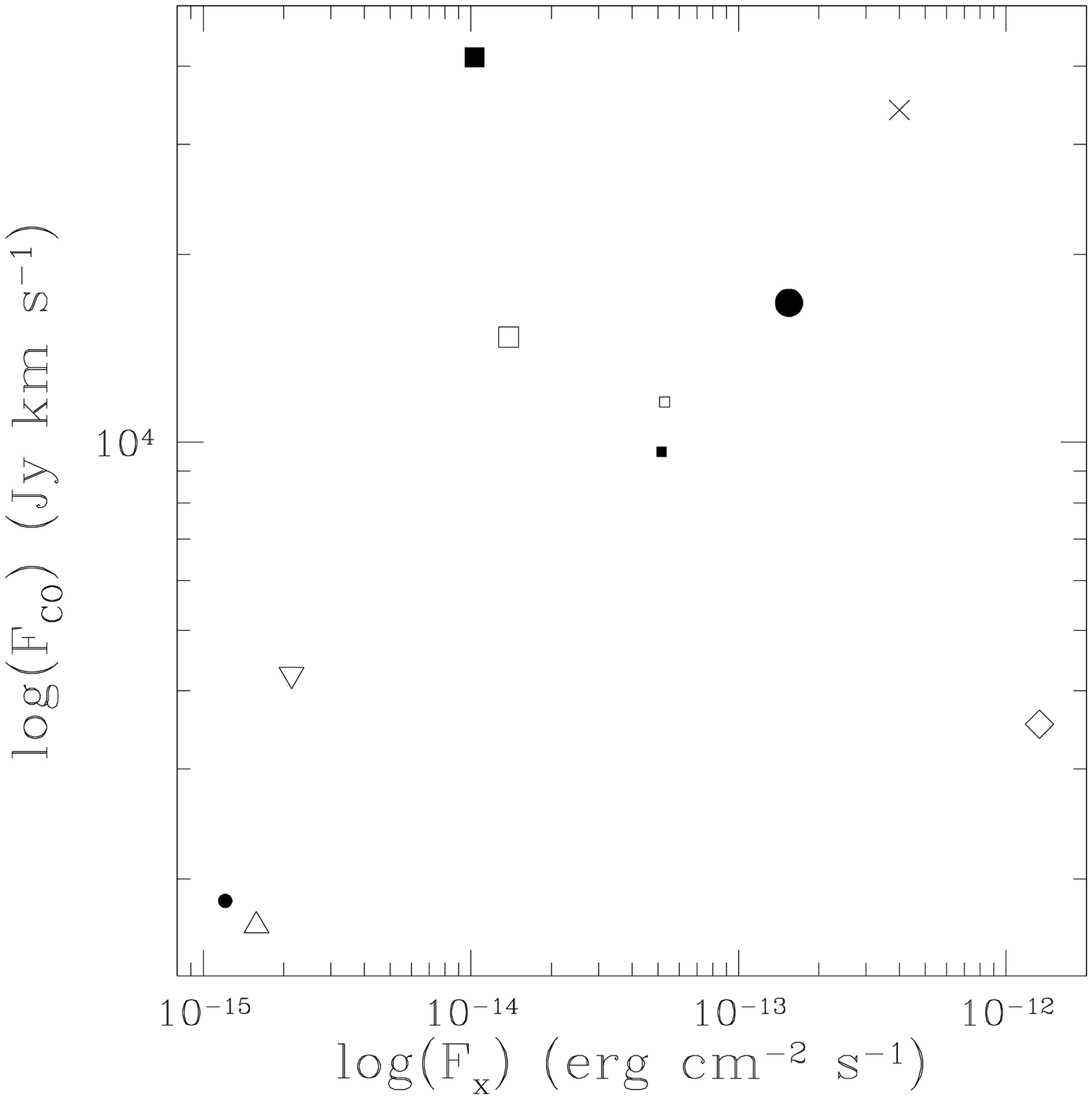}{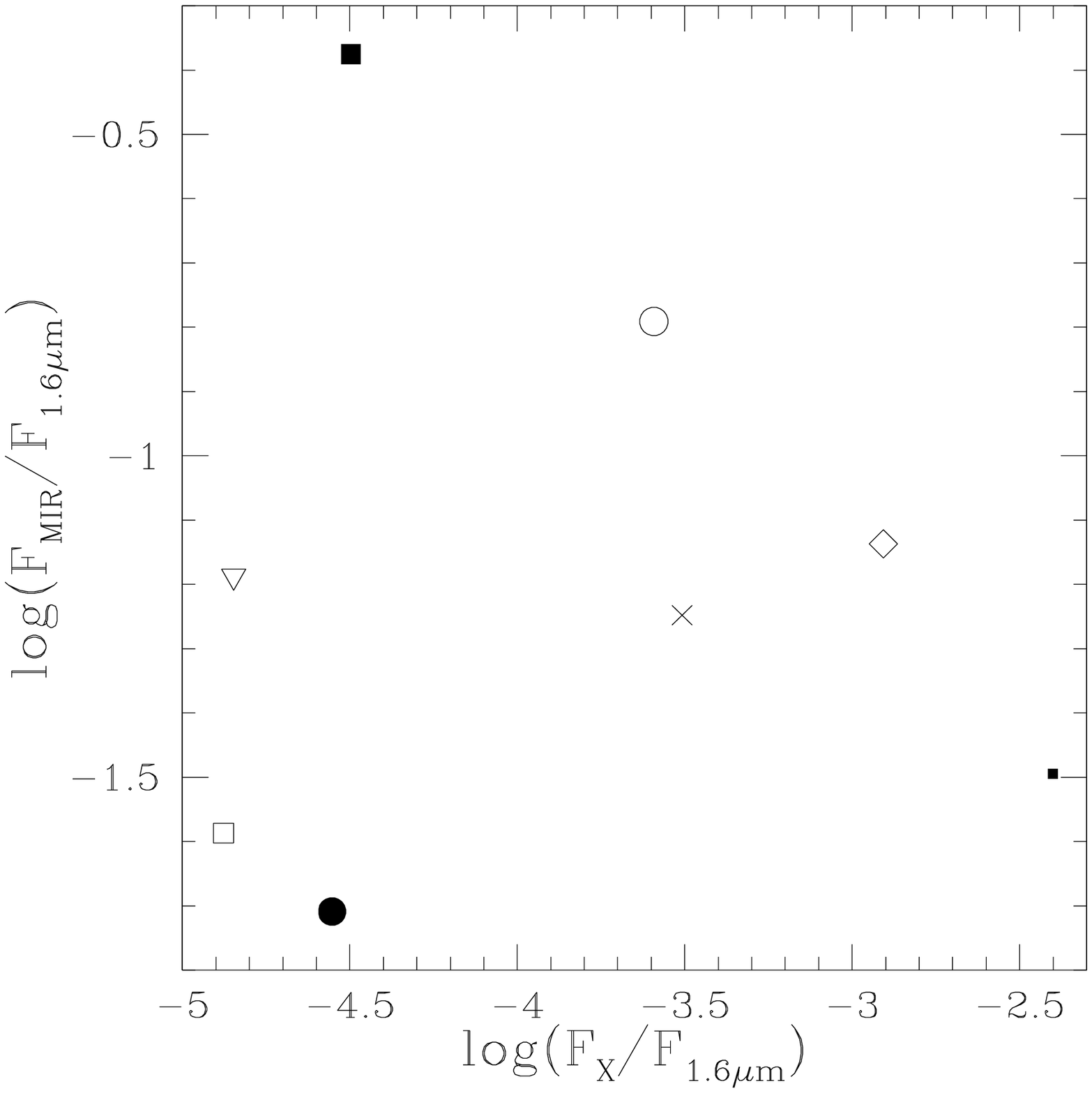}       
\caption[]{Continued.
c) A comparison between CO (from BIMA-SONG data) and X-ray fluxes in the 
centers of the galaxies, using the same apertures and data point shapes.
Again, the outlying points are NGC~6946 (upper left) and NGC~4579 (lower right).
d) Ratio of mid-IR and 1.6$\micron$ flux vs. ratio of X-ray and 1.6$\micron$ 
flux.  There is little correlation seen in this plot suggesting
that more than one factor determines the quantity of diffuse
X-ray emission in the centers of galaxies.
}
\end{figure*}

\clearpage 

\begin{figure*}
\epsscale{1.0}
\plotone{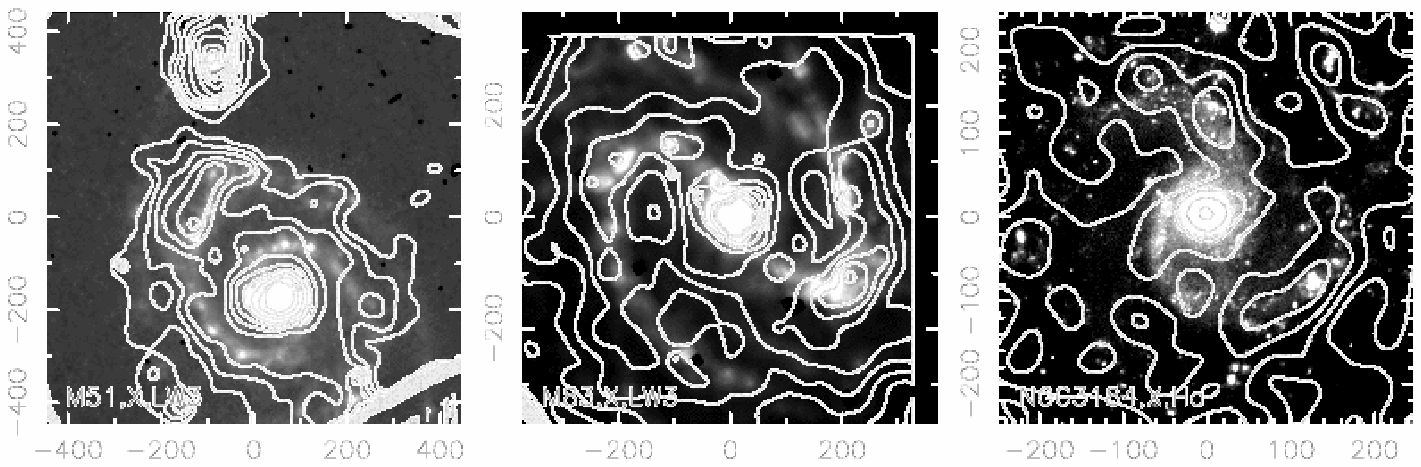}
\caption[]{
Diffuse X-ray contours on top of mid-infrared $15\micron$ emission (grayscale) 
for M51, M83 and H$\alpha$ (grayscale) for NGC~3184.
In the outer spiral arms of M51 and M83, some of the X-ray emission
extends to larger radius than the 
sites of star formation as seen in the mid-infrared image.
There is little evidence for this in NGC~3184 where the X-ray emission
lies centered on the regions of active star formation.
Population synthesis models predict a constant energy deposition
rate due to supernovae over a time period that is 10 times longer than
the lifetime of the massive stars associated with the HII regions
and so would have expected that the X-ray emission would significantly lag
the sites of active star formation.  The narrowness of the spiral
arms seen in the diffuse component and close association with
the site of active star formation suggest that interarm supernova
produce less soft X-ray emission than those near the site of
massive star formation.
\label{fig:overlay}
}
\end{figure*}
{}

\begin{deluxetable}{llccc}
\tablecolumns{3}
\tablecaption{Archived Nearby Galaxies 
with diffuse X-ray emission \label{table:intro}}
\tablehead{
\colhead{Galaxy}  &
\colhead{X-ray morphology}  &
\colhead{Observation ID }  &
\colhead{nuclear type }  &
\colhead{Distance (Mpc) } 
}
\startdata
 M51      & arms, center & 1622 & Sy2 & 7.7 \\
 M83      & arms, center & 793  & H2  & 4.5 \\
 M101     & arms, center & 934  &     & 7.4 \\
 NGC 2403 & irregular    & 2014 & H2  & 4.2 \\
 NGC 2681 & center       & 2060 & L1.9   & 13.3 \\
 NGC 3184 & arms, center & 804,1520 &H2& 8.7 \\
 NGC 4303 & arms, center & 2149 & H2  & 15.2 \\
 NGC 4579 & center       & 807  & S1.9/L1.9& 16.8 \\
 NGC 4736 & arms, center & 808  &  L2 & 4.3 \\
 NGC 5055 & center       & 2197 &  T2 & 7.2 \\
 NGC 6946 & arms,center  & 1043 &  H2 & 5.5 \\
 NGC 7331 & center       & 2198 &  T2 & 15.1 \\
\enddata
\tablecomments{
Observation IDs are those listed by the Chandra X-ray Center.
Nuclear types and distances are those compiled by \citet{helfer}
except for the distance to M83 which was measured by \citet{thim},
and that to NGC 2681 from \citet{Tully} assuming a Hubble constant
of 75 Mpc$^{-1}$ km s$^{-1}$.
}
\end{deluxetable}

\begin{deluxetable}{llccc}
\tablecolumns{3}
\tablewidth{0pt}
\tablecaption{X-ray and Mid-IR Fluxes in the Spiral Arms \label{table:arms}}
\tablehead{
\colhead{Galaxy}  &
\colhead{X-ray}  &
\colhead{Mid-IR}  \\
\colhead{ }  &
\colhead{(erg cm$^{-2}$ s$^{-1}$)} &
\colhead{(erg cm$^{-2}$ s$^{-1}$)}
}
\startdata
 M51      & 6.4e-15 & 1.8e-11 \\
          & 8.2e-15 & 1.0e-11 \\
          & 5.6e-15 & 1.7e-11 \\
          & 3.7e-15 & 8.7e-12 \\
          & 3.4e-15 & 4.8e-12 \\
          & 2.6e-15 & 1.5e-11 \\
 M83      & 1.6e-14 & 3.3e-11 \\
          & 1.0e-14 & 2.7e-11 \\
          & 6.0e-15 & 1.6e-11 \\
          & 4.0e-15 & 2.5e-11 \\
          & 7.2e-15 & 2.8e-11 \\
          & 5.6e-15 & 1.7e-11 \\
 M101     & 2.2e-15 & 4.7e-12 \\
          & 5.4e-16 & 2.7e-12 \\
          & 1.3e-15 & 6.1e-12 \\
          & 1.0e-16 & 3.0e-12 \\
          & 3.4e-16 & 3.0e-12 \\
          & 1.3e-16 & 2.1e-12 \\
          & 9.3e-16 & 6.2e-12 \\
 NGC 4579 & 5.5e-15 & 1.9e-12 \\
          & 2.8e-15 & 2.9e-12 \\
          & 3.6e-15 & 2.4e-12 \\
          & 8.8e-16 & 1.3e-12 \\
 NGC 4736 & 2.3e-14 & 1.8e-11 \\
          & 1.8e-14 & 1.8e-11 \\
          & 2.2e-14 & 1.6e-11 \\
          & 1.5e-14 & 2.8e-11 \\
          & 8.5e-15 & 2.0e-11 \\
 NGC 5055 & 1.8e-15 & 1.1e-11 \\
          & 1.9e-15 & 10.0e-12 \\
          & 3.0e-15 & 1.4e-11 \\
          & 3.0e-15 & 1.1e-11 \\
 NGC 6946 & 3.7e-15 & 9.9e-12 \\
          & 4.2e-15 & 8.3e-12 \\
          & 6.5e-15 & 5.9e-12 \\
          & 4.6e-15 & 1.4e-11 \\
          & 3.6e-15 & 9.5e-12 \\
\enddata
\tablecomments{
Fluxes are measured in $20''$ apertures in the spiral 
arms of given galaxies. X-ray fluxes
have been corrected for Galactic absorption.  
Both X-ray and mid-infrared fluxes 
have been background-subtracted.  
Mid-infrared fluxes have been measured from ISOCAM 12 or 15$\mu$m
images using $\nu F_\nu$ and calibration factors given by
the ISO archive.
A plot of these data is shown in Figure \ref{fig:mir_all}.}
\end{deluxetable}

\begin{deluxetable}{lcccccc}
\tablecolumns{6}
\tablecaption{X-ray, Mid-IR, CO, and 2MASS IR Core Measurements 
\label{table:core}}
\tablehead{
\colhead{Galaxy  }  &
\colhead{X-ray}  &
\colhead{Mid-IR}  &
\colhead{CO}  &
\colhead{$H$-band}  &
\colhead{Aperture}  &
\colhead{Galactic $N_{H}$ Column}  \\
\colhead{ }  &
\colhead{(erg cm$^{-2}$ s$^{-1}$)}  &
\colhead{(erg cm$^{-2}$ s$^{-1}$)}  &
\colhead{(Jy km s$^{-1}$)}  &
\colhead{(mag)}  &
\colhead{($''$)}  &
\colhead{$10^{20}{\rm cm}^{-2}$}
}
\startdata
 M51      & 4.0e-13 & 9.1e-11 & 3.4e4  & 7.7  & 25 & 1.52 \\
 M83      & 5.8e-13 & 3.6e-10 & \nodata& 7.3  & 20 & 3.70 \\
 M101     & 2.1e-15 & 9.8e-12 & 4.2e3  & 10.2 & 10 & 1.15 \\
 NGC 2403 & 1.6e-15 & \nodata & 1.7e3  & 9.4  & 20 & 4.17 \\
 NGC 2681 & 4.9e-15 & \nodata & \nodata& 8.8  & 10 & 2.48 \\
 NGC 3184 & 1.2e-15 & \nodata & 1.8e3  & 10.9 & 10 & 1.15 \\
 NGC 4303 & 5.3e-14 & \nodata & 1.2e4  & 9.4  & 10 & 1.65 \\
 NGC 4579 & 1.3e-12 & 7.8e-11 & 3.5e3  & 8.1  & 20 & 2.51 \\
 NGC 4736 & 1.5e-13 & 1.1e-10 & 1.7e4  & 6.4  & 20 & 1.44 \\
 NGC 5055 & 1.4e-14 & 2.7e-11 & 1.5e4  & 8.1  & 15 & 1.30 \\
 NGC 6946 & 1.0e-14 & 7.7e-10 & 4.1e4  & 9.4  & 10 & 20.05 \\
 NGC 7331 & 5.2e-14 & 4.1e-11 & 9.7e3  & 7.6  & 20 & 8.25 \\
\enddata
\tablecomments{Values in the cores of galaxies for the given apertures for 
the diffuse X-ray, Mid-IR, CO, and 2MASS $H$-band images. X-ray fluxes have 
been corrected for the indicated Galactic absorption.  X-ray, mid-IR, 
CO, and 2MASS IR have all been background-subtracted.  These data points
are plotted in Figure \ref{fig:x.core}.
}
\end{deluxetable}

\end{document}